\documentclass[amsmath,12pt,amssymb,preprint,nofootinbib]{revtex4}
\usepackage{amsfonts} 
\usepackage{graphicx} 
\usepackage{epsfig}
\usepackage{multirow}
\usepackage{bm}

\begin{document}

\title{\textbf{Open Bottom mesons in magnetized matter- effects of (inverse) magnetic catalysis}}
\author{Sourodeep De}
\email{sourodeepde2015@gmail.com}
\author{Pallabi Parui}
\email{pallabiparui123@gmail.com}
\author{Amruta Mishra}
\email{amruta@physics.iitd.ac.in}  
\affiliation{Department of Physics, 
Indian Institute of Technology, Delhi, Hauz Khas, New Delhi - 110016}
\begin{abstract}
In-medium masses of the pseudoscalar and vector open bottom mesons ($B$, $\bar{B}$, $B_s$ and $B^*$, $\bar{B}^*$, $B_s^*$) are studied in the magnetized nuclear matter by considering the effects of Dirac sea, within the chiral effective model. The mass modifications arise due to the interactions of the open bottom mesons with the nucleons and the scalar mesons, calculated in terms of the scalar and number densities of the nucleons and the scalar fields fluctuations. The effects of the magnetized Dirac sea lead to the considerable changes in the scalar fields with magnetic field, which are related to the condensates of light quark–antiquark pairs. There is observed to be a (reduction) enhancement in the light quark condensates with magnetic field, a phenomenon called (inverse) magnetic catalysis. The contribution of the magnetic field on the Fermi sea of nucleons are taken into account through the Landau energy levels of protons and anomalous magnetic moments (AMMs) of the nucleons. The additional contribution of the lowest Landau level for the charged mesons are considered. The spin-magnetic field interaction between the longitudinal component of the vector and the pseudoscalar mesons ($B^{*||}(\bar{B}^{*||})-B (\bar{B}))$, and ($B_s^{*||}-B_s$) are studied, which lead to a level repulsion between their masses with magnetic field. Magnetic fields are observed to have significant contribution on the in-medium masses of the open bottom mesons through the Dirac sea effect as compared to the case when this effect is not considered. In vacuum, considerable changes are
obtained only due to the magnetized Dirac sea at zero and finite nucleonic AMMs.

\end{abstract}
\maketitle

\section{Introduction}
\label{sec1}
The study of the in-medium properties of heavy flavor hadrons \cite{Hosaka} has been an interesting topic of research in the studies of the hadron properties in extreme conditions. This topic has intense research interest due to its relevance in the high energy heavy ion collision experiments, where matter at very high temperature and/or density can be created from the collisions of heavy ions at the ultra relativistic high energy. In the non-central heavy ion collision experiments, magnetic fields of  large magnitudes have been estimated in many works \cite{kharzeev, fukushima, skokov, deng, tuchin} ($eB\sim 2m_{\pi}^2$ at RHIC in BNL and $eB\sim 15m_{\pi}^2$ at LHC in CERN). The time evolution of the produced field is still an open question, needs to solve the magnetohydrodynamic equations, with the proper estimation of the electrical conductivity of the medium \cite{tuchin, tuchin1, tuchin2, arpan}. The magnetic field produced is weak in the (near) central collisions due to the small impact parameter of collision and the produced medium becomes dense. On the contrary, in the non-central heavy ion collisions, produced magnetic fields can be very large accompanied by low density medium. The effects of magnetic field on the heavy flavor mesons should have observable consequences in the high energy heavy ion collision experiments since they are formed in the early stages of collisions, when the produced magnetic field can still be large. 

There have been a lot of studies in the literature, on the in-medium properties of the heavy flavor mesons (both the heavy quarkonia and the open heavy flavor mesons) in the presence and absence of magnetic field, using the various approaches of chiral effective model \cite{sushruth, nikhil, amal98, amalarxiv1, arindam, am47, div90, bottomstrange, div91, am81}, QCD sum rule approach \cite{klingl, kim, ko, am82, pallabi, pallabibottom, morita1, morita2, arata, hilger, hilger1, wang, matheus}, potential models \cite{eichten1, eichten2, radfort}, the coupled channel approach \cite{molina, tolos}, Quark meson coupling (QMC) model \cite{krein, tsushima}, heavy meson effective theory \cite{sudoh}, etc. The studies in the QMC model show attractive interactions of the $J/\psi$ as well as of the open heavy flavor mesons ($\bar{D}, B$) in nuclear matter, suggesting the possibility of creating meson-nuclei bound states \cite{krein}. The in-medium masses of the heavy flavor mesons have been investigated within a chiral effective model. The chiral effective model in its original $SU(3)$ version \cite{papa59}, is generalized to incorporate the interactions of the heavy flavor (charm and bottom) hadrons with the baryons and scalar mesons, in presence of a hadronic medium. The masses of the heavy quarkonia (charmonia and bottomonia) are obtained from the medium modifications of a scalar dilaton field, $\chi$ which corresponds to the QCD gluon condensates \cite{div90, am47, am81}. 
The effects of the magnetic field have been studied on the masses of the open charm \cite{sushruth, 102} and open bottom mesons \cite{nikhil, amsm31} in magnetized nuclear medium within the chiral effective model, considering the magnetic field effects through the Landau quantization of protons and AMMs of the nucleons. The magnetic
field contribution is through the Landau energy levels of the protons and the
AMMs of the nucleons via the scalar and the number densities of the protons and
neutrons in the nuclear matter \cite{nikhil, sushruth, prakash, broderik, wei, guang}. The charged heavy flavor mesons ($D^{\pm}$ and $B^{\pm}$) have additional mass shifts due to the lowest Landau level (LLL) contribution at finite magnetic field \cite{nikhil, sushruth, amsm31, amsm30}. In the QCD sum rule approach, the in-medium masses of the $1S$ and $1P$ wave states of charmonium and bottomonium have been calculated in the magnetized nuclear matter, from the medium modifications of the scalar and twist-2 gluon condensates calculated within a chiral SU(3) model \cite{pallabi, pallabibottom}. In ref.\cite{mc1}, the Dirac sea contribution in presence of an external magnetic field have been studied on the heavy quarkonia masses in magnetized nuclear matter by using the QCD sum rule approach. The changes in the masses are appreciably visible in this case with variation in magnetic field. 

The PV mixing  between the longitudinal component of the vector and the pseudoscalar charm (bottom) mesons at finite magnetic field, has been studied in many works \cite{amsm30, amsm31, machado, gubler, cho91, 102, suzuki, iwasaki, mc1, pallabi, pallabibottom} and the magnetic fields are seen to have dominant contribution through the PV mixing effects on their masses as well as decay widths. The in-medium hadronic decays of the vector open charm meson, $D^*\rightarrow D\pi$ have also been studied incorporating the PV mixing effects between $D-D^{*}$ and LLL contributions for the charged mesons at finite magnetic field \cite{amsm30}. The (decrement) increment of the QCD light quark condensates with increasing magnetic field is known as (inverse) magnetic catalysis \cite{elia, kharmc1, chernodub, kharzeevmc, gui}. As the temperature and density dependence of the QCD condensates can modify the hadron properties, the magnetic field modifications of the condensates can contribute significantly on the hadronic properties. In ref.\cite{yoshida}, the contribution of (inverse) MC has been studied on the neutral open heavy flavor mesons through the change in the light constituent quark mass with magnetic field. In the literature, there have been studies of the magnetized Dirac sea effects on the quark matter properties within the Nambu-Jona-Lasinio (NJL) model \cite{Preis, menezes, ammc, lemmer, guinjl}. In ref.\cite{haber}, effects of magnetized Dirac sea have been studied on the nuclear matter phase transition using the Walecka model and the extended linear sigma model. The work shows a rise in the nucleon mass with increasing magnetic field at zero density and zero anomalous magnetic moments of the nucleons, indicating the magnetic catalysis effect indirectly through the scalar field dependency of the nucleon mass. The authors of \cite{arghya}, have been studied the effects of magnetized Dirac sea in the weak-field approximation, by evaluating the modified fermion propagator in the magnetic field following the nucleon self-energy calculation through summation of the nucleonic tadpole diagrams. An increase of nucleon mass with magnetic field was observed in the Walecka model at zero density, with significant contribution from the anomalous magnetic moments of the Dirac sea of nucleons. At finite temperature, decrease of the critical temperature of vacuum to nuclear matter phase transition with magnetic field, indicates an inverse magnetic catalysis (IMC) \cite{balicm} for nonzero nucleonic AMMs, whereas an opposite behavior is obtained for zero AMM of the nucleons.\cite{arghya} 

In our present study, we have introduced the effects of (inverse) magnetic catalysis on the in-medium properties of open bottom mesons, due to the incorporation of the magnetized Dirac sea effects in the chiral model framework. Thus, the in-medium masses of the pseudoscalar $B^+ (\bar{b}u$), $B^- (\bar{u}b)$, $B^0 (\bar{b}d)$, $\bar{B}^0(\bar{d}b)$, $B^0_s(\bar{b}s)$, $\bar{B}^0_s(\bar{s}b)$ and the vector open bottom mesons $B^{*+}$, $B^{*-}$, $B^{*0}$, $\bar{B}^{*0}$, $B^{*0}_s$, $\bar{B}^{*0}_s$ (with the same quark-antiquark constituents as their pseudoscalar partners) are studied in the magnetized nuclear matter, accounting for the Dirac sea effects. The contribution of the lowest Landau level for the charged mesons are also considered at finite magnetic field. The level repulsion due to the spin-magnetic field interaction effects on the pseudoscalar and longitudinal component of vector open bottom mesons are studied in presence of Dirac sea contribution.  

The present paper is organized as follows, in section \ref{sec2}, the in-medium masses of the open bottom mesons in magnetized nuclear matter are described using the chiral effective model. The sub-section \ref{secA}, illustrates the interaction Hamiltonian to find the spin-mixing effects between the longitudinal component of the spin-triplet (vector) and the spin-singlet (pseudoscalar) meson states, incorporating the Dirac sea effects in presence of an external magnetic field. The results of the present study are discussed in section \ref{sec3}. Section \ref{sec4}, presents the summary of the present investigation.  

\section{In-medium masses of the open bottom mesons}
\label{sec2}
The in-medium masses of the open bottom mesons, $B (B^+, B^0)$ and $\bar{B} (B^-, \bar{B}^0)$ are studied within a chiral effective model Lagrangian by generalizing the chiral $SU(3)$ model to incorporate the interactions between the open heavy flavor (bottom) mesons and the light hadrons \cite{nikhil, amsm31, div91, amalbottom}. The original chiral $SU(3)_L\times SU(3)_R$ model with three light quark flavors, is based on the non-linear realization of chiral symmetry \cite{coleman, weinberg, bardeen} and the QCD broken scale invariance \cite{papa59, zschii}. A logarithmic potential in terms of a scalar dilaton field, $\chi$ \cite{erikk} represents the scale symmetry breaking effect of QCD in the chiral model. The general form of the Lagrangian density is written as \cite{papa59}
\begin{equation}
    \mathcal{L} = \mathcal{L}_{kin} + \sum_{W} \mathcal{L}_{BW} + \mathcal{L}_{vec} + \mathcal{L}_{0} +\mathcal{L}_{SB} + \mathcal{L}_{scale break} + \mathcal{L}_{mag}
\end{equation}
where, $\mathcal{L}_{kin}$ represent the kinetic energy terms of baryons and mesons; $\mathcal{L}_{BW}$ gives the baryon-meson interactions for $W=$ spin-0 (scalar, pseudoscalar) and spin-1 (vector, axial-vector) mesons; $\mathcal{L}_{vec}$ contains the dynamical mass generation of the vector mesons mass through the interactions with the scalar fields and the vector mesons quartic self-interactions terms;  $\mathcal{L}_{0}$ corresponds to the meson-meson interactions; $\mathcal{L}_{SB}$ is the explicit chiral symmetry breaking term; $\mathcal{L}_{scale break}$ corresponds to a scale invariance breaking logarithmic potential given in terms of the scalar dilaton field, $\chi$. Finally, the magnetic term, $\mathcal{L}_{mag}$ gives the interaction of the baryons with an external electromagnetic field, of field strength tensor $F^{\mu\nu}$. This last term includes apart from the photon kinetic energy, baryon photon interactions through the magnetic vector potential, a tensorial interaction term of the baryon and electromagnetic field which incorporates the effects of the anomalous magnetic moments of the baryons in presence of a finite magnetic field \cite{nikhil, sushruth, amal98}. This incorporate the effects of magnetic field due to the discretized Landau energy levels of protons and anomalous magnetic moments of the nucleons in the number ($\rho_i; i=p,n$) and scalar densities ($\rho_i^s$) of nucleons in the magnetized nuclear matter \cite{erikk, strikland, broderik, prakash, wei, guang, ivanov}  The Euler Lagrange's equations motion are solved for the scalar isoscalar fields, $\sigma$ (non-strange), $\zeta$ (strange), the scalar isovector field, $\delta$ (non-strange) and the scalar dilaton field, $\chi$. The scalar mesons are treated as classical fields and the nucleons as the quantum fields, to incorporate the additional effect of the Dirac sea on the equations of motion of the scalar fields, through the scalar densities of the nucleons. In this approximation, the scalar fields, which treated as classical fields, are related to the QCD condensates $\big[\sigma (\sim (\langle \bar{u}u \rangle + \langle \bar{d}d \rangle))$, $\zeta (\sim \langle \bar{s}s \rangle)$ and $\delta(\sim (\langle\bar{u}u \rangle -\langle\bar{d}d \rangle))\big]$. 

In our present work, we have studied the in-medium masses of the open bottom mesons in the magnetized, asymmetric nuclear matter, accounting for the effects of the Dirac sea. At the given values of baryon density, $\rho_B$, magnetic field, $|eB|$ (in units of $m_{\pi}^2$) and isospin asymmetry parameter, $\eta = \frac{\rho_n-\rho_p}{2\rho_B}$ (where $\rho_n$ and $\rho_p$ are the number densities of the neutron and proton, respectively) of the nuclear medium, the coupled equations of motion in the scalar fields are solved. The contributions of magnetic fields on the Fermi sea of nucleons in the magnetized (nuclear) matter, are incorporated through the scalar and number densities of protons ($\rho^s_p, \rho_p$) and neutrons ($\rho_n^s, \rho_n$), due to the Landau energy levels of the charged protons and the anomalous magnetic moments of the nucleons \cite{wei, nikhil, sushruth, amal98, prakash}. The Dirac sea contributed on the scalar densities of the nucleons and hence on the scalar fields in the chiral effective model. The contribution is obtained through the summation over the scalar ($\sigma$, $\zeta$ and $\delta$) and vector ($\rho$ and $\omega$) tadpole diagrams in the one-loop self energy calculation of nucleons, in the weak-field expansion of nucleonic propagators at finite magnetic field \cite{arghya}.\\
To obtain the in-medium masses of the open bottom mesons ($B$ and $\bar{B}$), the dispersion relations as obtained from the Fourier transformations of the equations of motion of these mesons from the chiral effective Lagrangian \cite{nikhil, amsm31, div91,bottomstrange}, are solved
\begin{equation}
    - \omega^2 + \vec{k}^2 + m^2_{B({\bar{B}})} - \Pi_{B(\bar{B})}(\omega, |\vec{k}|) = 0
    \label{dis}
\end{equation}
Where $\Pi_{B(\bar{B})}(\omega, |\vec{k}|) $ represent the self-energy functions of the $B (\bar{B})$ mesons doublet [$B(B^+, B^0)$ and $\bar{B} (B^-, \bar{B}^0)$] in magnetized nuclear medium. The expressions for the self-energy functions of the $B (\bar{B})$ mesons doublet are given in terms of the proton and neutron number densities ($\rho_p,   \rho_n$) and scalar densities ($\rho^s_p, \rho^s_n$) in the magnetized nuclear matter; some fitted parameters of the effective Lagrangian (the B meson decay constant, $f_B$ and two other parameters, $d_1, d_2$ fitted from the low-energy kaon-nucleon scattering lengths for isospin $I=0$ and $I=1$ channels \cite{div91}), and the fluctuations of the scalar fields, $\sigma'(=\sigma-\sigma_0)$, $\zeta'_b(=\zeta_b-\zeta_{b0})$ and $\delta' (=\delta-\delta_0)$ from their vacuum expectation values ($\sigma_0, \zeta_{b0}, \delta_0$). Due to the small fluctuations of the heavy quark condensates within the medium, the fluctuation of $\zeta_b  \ (\sim \langle\bar{b}b\rangle$), is neglected in our work. The masses of the bottom strange mesons are obtained by solving the similar dispersion relation for $\omega$ at $\vec{k}=0$, as given by eq. (\ref{dis}) with the corresponding parameters for the $B_s^{0}$ and $\bar{B}^{0}_{s}$ mesons. The self-energy functions for the particle and antiparticle of $B_s$ meson have similar expressions in nuclear matter \cite{bottomstrange}, and hence their in-medium masses are the same in nuclear matter. In this case, the medium modifications through the scalar field's fluctuation are obtained due to the strange, isoscalar field $\zeta$ ($\sim \langle\bar{s}s\rangle$). 
\begin{equation}
    \Pi(\omega,|\textbf{k}|)_{B_s} = \Bigg[\frac{d_1}{2f_{B_s}^2}(\rho_p^s+\rho_n^s)-\frac{\sqrt{2}}{f_{B_s}}(\zeta'+\zeta_b')\Bigg](\omega^2-\vec{k^2}) + \frac{m_{B_s}^2}{\sqrt{2}f_{B_s}}(\zeta'+\zeta_b').
    \label{selfBs}
\end{equation}
The medium modifications of others scalar fields are obtained from the coupled equations of motion at given values of the baryon density ($\rho_B$), isospin asymmetry parameter ($\eta$) of the nuclear medium in presence of an external magnetic field ($|eB|$). In our present work, the effects of the nucleons' anomalous magnetic moments are considered in the Fermi sea via the tensorial interaction term in $\mathcal{L}_{mag}$ and also on the magnetized Dirac sea in terms of the weak-field expansion of the fermionic propagators, which are obtained by solving the Dirac equation taking into account its interaction with an external magnetic field due to electric charge and non zero anomalous magnetic moments of Dirac sea nucleons.\\
The charged pseudoscalar $B^{\pm}$ mesons have additional contribution from the Landau energy levels in the presence of magnetic field. Considering the contribution from the lowest Landau level ($n=0$) only, the effective masses of the charged mesons are given by \cite{chernodubll, taya, amsm31} 
\begin{equation}
    m^{eff}_{B^{\pm}}=\sqrt{m_{B^{\pm}}^{*2}+ |eB|}
\end{equation}
This formula refers to the contribution of the lowest Landau level on a charged pseudoscalar particle, ignoring its internal structure \cite{chernodubll, amsm31}. The masses of the neutral mesons ($B^0, \bar{B}^0$) have no such Landau level contribution due to their charge neutrality. 
The in-medium masses of $m^*_{B^{\pm}, B^0, \bar{B}^0}$ are calculated within the chiral effective model by solving their dispersion relations for $\omega$ at $\vec{k}=0$, as given by equation (2).\\
The masses of the vector open bottom mesons $B^*$ ($B^{*+}$, $B^{*0}$) and $\bar{B}^*$ ( $B^{*-}$, $\bar{B}^{*0}$) mesons, which have the same quark-antiquark constituents as $B$ and  $\bar{B}$ mesons, are assumed to have identical mass shifts as
the shifts in the masses of the $B$ and  $\bar{B}$ mesons calculated within the chiral effective model. This
is in line with the mass modifications of hadrons within the Quark meson coupling (QMC) model, which arise due to the
modification of the scalar density of the light quark (antiquark) constituent of the hadron \cite{Hosaka}. Thus the mass shifts of the $B^{*}$ and $\bar{B}^*$ mesons, in the magnetized nuclear matter, are assumed to be \cite{amsm31}
\begin{equation}
    m^*_{{B^*},(\bar{B}^*)} - m^{vac}_{{B^*},(\bar{B}^*)} = m^*_{{B},(\bar{B})} - m^{vac}_{{B},(\bar{B})}
    \label{diff}
\end{equation}

For the charged $B^{*\pm}$ mesons, contribution from the lowest Landau level (LLL) ($n=0$) contribution on the masses obtained from equation (5) \cite{chernodubll, amsm31}
\begin{equation}
    m^{eff}_{{B^{*\pm}}} = \sqrt{{m_{B^{*\pm}}^{*}}^2+ (-gS_z+1)|eB|}
    \label{lllvec}
\end{equation}
As it is noted from equation (6), the effective masses of the charged mesons depend on the spin-projection along the direction of the external magnetic field, $\vec{B}=B\hat{z}$, namely on the $S_z$ component of the intrinsic spin of the vector bottom mesons with $S=1$. For transverse component with $S_z=-1$, square of the vector particle mass increases by $3|eB|$ amount and of the $S_z=1$ component decreases by $|eB|$ amount, with magnetic field for the gyromagnetic ratio $g=2$ \cite{amsm30, amsm31, chernodubll}. However, in the presence of an external magnetic field, there is spin-mixing between the longitudinal component ($S_z=0$) of the vector mesons $B^{*}$ ($\bar{B}^*$) and the pseudoscalar meson $B$ ($\bar{B}$). This is described in the next subsection. Therefore, we consider only the effective mass of the longitudinal component ($S_z=0$) of the charged $B^{*\pm}$ mesons, to be used in the PV mixing calculation \cite{amsm31}
\begin{equation}
   m^{eff}_{{B^{*\pm}}^{(||)}}=\sqrt{{m^*_{B^{*\pm}}}^2 + |eB|}
   \label{lllsz}
\end{equation}
In equations (\ref{lllsz}), the in-medium masses $m^*_{B^*, \bar{B}^*}$ of the vector mesons are obtained from the in-medium masses of the corresponding pseudoscalar partners [using equation (\ref{diff})], as calculated within the chiral effective model framework.

\subsection{Interaction Hamilton for the Pseudoscalar-Vector Mesons (PV) Mixing}
\label{secA}
In this sub-section, we present the formulation to evaluate the mass modifications of the pseudoscalar and the longitudinal component of the vector meson due to the mixing effect between these states in the presence of magnetic field. The Hamilton accounting for the spin-magnetic field interaction is given by \cite{amsm31, alford, iwasaki, yoshida}
\begin{equation}
    H_{spin-mix.} = -\sum_{i=1}^{2} \vec{\mu}_{i}.\vec{B} 
    \label{hamil}
\end{equation}
where, $\vec{\mu_i}=gq_i \vec{S_i}/2m_i$ is the quark magnetic moment for the $i^{th}$ flavor, presented in the bound states of open bottom mesons ($\bar{q_2}q_1$). In this equation, $g$ is the Lande g-factor, is taken to be 2. $q_i$ is the electric charge (in units of the electron charge $|e|$), $\vec{S_i}$ denotes the spin and $m_i$ is the mass of the $i^{th}$ flavor of quark. Solving the eigenvalue problem for the effective Hamiltonian (\ref{hamil}) by considering the two-dimensional eigensystem of spin-singlet ($|00\rangle$) and spin-triplet ($|10\rangle$) states, leads to a rise and drop in the mass of the $S_z=0$ component of the vector and the pseudoscalar mesons, respectively, in the following way 
\begin{equation}
    m^{PV}_{V^{||}} = m^{eff}_V + \Delta m_{sB} \quad m^{PV}_{P} = m^{eff}_P - \Delta m_{sB};
\end{equation}
This effect is also called the level repulsion with $ \Delta m_{sB} = \frac{\Delta E}{2} ((1+x^2)^{1/2}-1)$;  $x = \frac{2}{\Delta E}\frac{(-g|eB|)}{4}(\frac{q_1}{m_1}-\frac{q_2}{m_2})$; $\Delta E = m^{eff}_V - m^{eff}_P$ is the mass difference between the vector and pseudoscalar mesons calculated within the chiral effective model. In this work, we will be studying the spin-mixing effects between the ($B-B^*$), ($B_s-B_s^*$), ($\bar{B}-\bar{B}^*$) and ($\bar{B}_s-\bar{B}_s^*$) mesons accounting for the additional effects of the (inverse) magnetic catalysis on their masses, in magnetized nuclear matter due to the contribution from the magnetized Dirac sea.

\section{Results and Discussion}
\label{sec3}
In this work, the in-medium masses of the pseudoscalar, $B^{\pm}$, $B^0$, $\bar{B}^0$,  the vector, $B^{*\pm}$, $B^{*0}$, $\bar{B}^{*0}$ open bottom mesons as well as the bottom strange mesons $B_s^0$, $\bar{B}_s^0$, $B_s^{*0}$, $\bar{B}_s^{*0}$, are studied in the magnetized asymmetric nuclear matter, accounting for the effects of Dirac sea (denoted as 'DS'), using the generalized version of the chiral effective model. The mass modifications of the mesons arise due to their interactions with the scalar mesons and nucleons in nuclear medium. The in-medium masses are obtained by solving equation (2) for the pseudoscalar open bottom mesons, with the self-energies for all $B^{+}$, $B^{-}$, $B^0$, $\bar{B}^0$, $B_s^0$ and $\bar{B}_s^0$ meson states calculated in terms of the number and scalar densities of the nucleons and the scalar fields fluctuations with respect to their vacuum expectation values \cite{nikhil, div91, bottomstrange}. The scalar and number densities ($\rho^s_i$ and $\rho_i;\ i=p,n$) of nucleons receive magnetic field contributions due to the Landau energy levels of the positively charged protons and anomalous magnetic moments (AMMs) of the nucleons in the magnetized nuclear matter. In our present study, the effects of Dirac sea at finite magnetic fields are taken into account while calculating the in-medium mass of the open bottom mesons. Dirac sea contributions are incorporated through the nucleonic tadpole diagrams corresponding to the various meson-nucleon interactions of scalar ($\sigma$, $\zeta$ and $\delta$) and vector ($\rho$, and $\omega$) meson particles. The one-loop self energy functions of the nucleons are obtained in the weak-field approximation of the fermionic propagator by assuming terms up to 2nd order in $B$ and nucleons' anomalous magnetic moments. The Dirac sea effect, thus, contributes to the scalar densities of the nucleons in the chiral model framework. The scalar fields are modified due to the magnetized Dirac sea while solving their coupled equations of motion under the classical fields assumptions from chiral model Lagrangian density. The magnitudes of the scalar fields, thus solved, are observed to increase with increasing magnetic field at zero baryon density $(\rho_B=0)$, for non zero as well as zero values of the anomalous magnetic moments (AMM) of the nucleons, but the solutions of the fields for non-zero AMMs of nucleons are obtained till $|eB|\approx 3.9m_{\pi}^2$, in our current study. For e.g., the values of the scalar fields $\sigma$ and $\zeta$ (in MeV) at $\rho_B=0$ and $|eB|=0$, of $-93.3$ and $-106.7$ are modified to $-96.4\ (-93.8) $ and $-107.8\ (-106.9)$ for $|eB|=2 m_\pi^2$, for non zero (zero) AMMs of the nucleons. These fields correspond to the QCD light quark condensates as described in section (\ref{sec2}). The enhancement of the light quark condensates with rising magnetic field is recognised as magnetic catalysis. In the vacuum, in presence of an external magnetic field, there is no contribution from the protons' Landau energy levels in the absence of the matter part. For the presence of a light quark flavor ($u$, $d$ or $s$) in the $B$ mesons' constituent quark structure, there are observed to be significant changes in their masses due to the change in the light quark condensates with magnetic field, accounting for the effects of Dirac sea. The anomalous magnetic moments of the protons and the neutrons are found to have considerable effects through the Dirac sea effect via magnetized fermionic propagators. At the nuclear matter saturation density, $\rho_0$, both in the symmetric ($\eta=0$) as well as the asymmetric ($\eta=0.5$) nuclear matter, the magnitudes of the scalar fields are observed to decrease with magnetic field for non zero AMMs of the nucleons, indicating an inverse magnetic catalysis (IMC), as they are proportional to the light quark condensates. For the case of zero AMM in Dirac sea, at $\rho_0$, opposite behavior is observed, i.e., scalar fields tend to rise with magnetic field, leading to the magnetic catalysis (MC). The effects of DS on the scalar fields and hence on the light quark condensates are observed to be much more pronounced with changing magnetic field, than the situation of only considering the magnetized Fermi sea contribution through the protons' Landau quantization. Thus, there are observed to be appreciable effects of the magnetic field on the open bottom meson masses, accounting for the Dirac sea contribution, in presence of an external magnetic field. The nuclear matter saturation density is taken to be $\rho_0=0.15 \ fm^{-3}$ in our work. The constituent quark masses (in MeV) are taken to be $m_u=m_d=330$ MeV and $m_b=5360$ in the calculation of the spin-magnetic field interaction effects. 
\begin{figure}[h!]
    \centering
\includegraphics[width=1.0\textwidth]{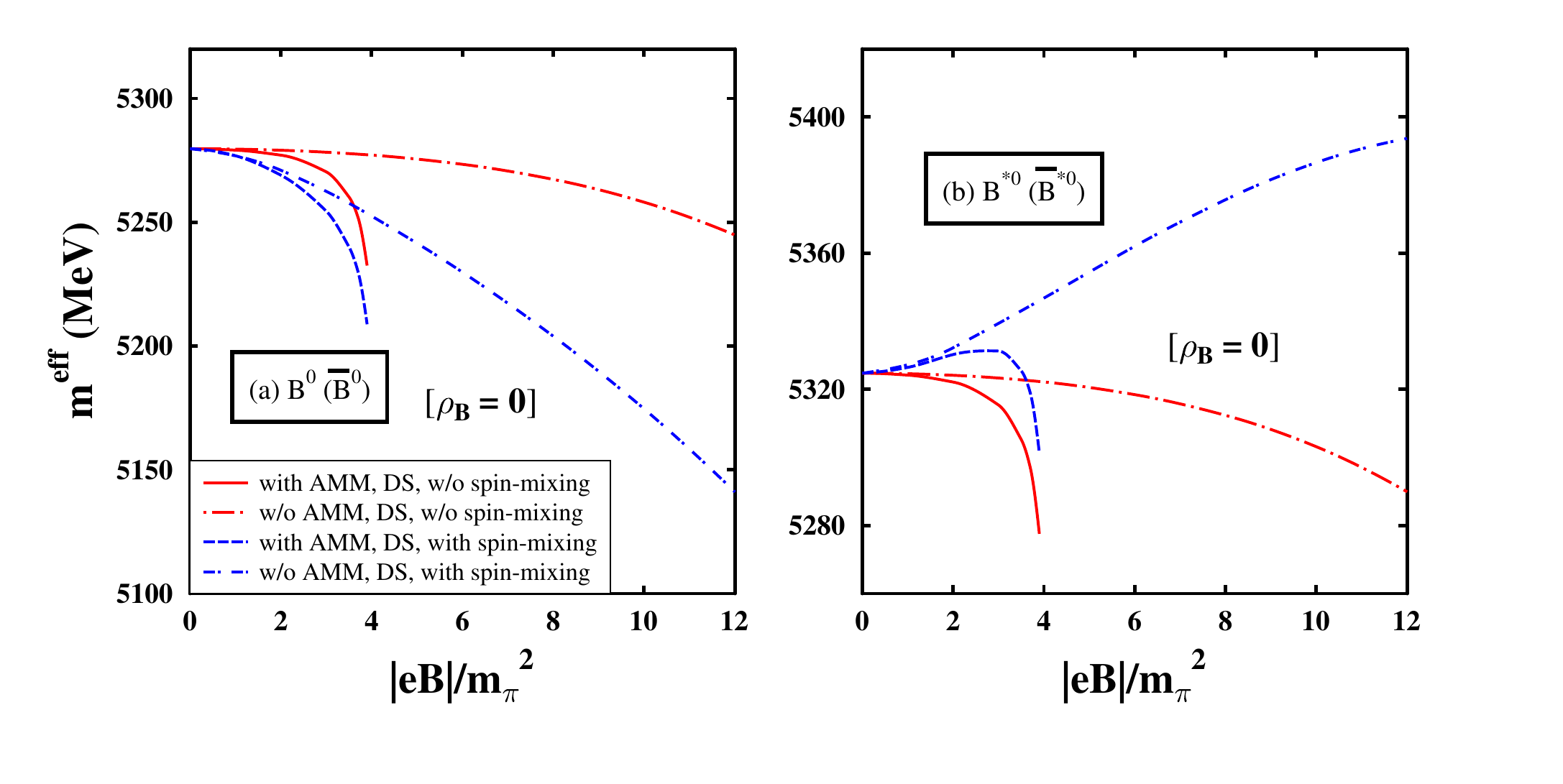}
\vspace{-2.5em}
    \caption{In-medium masses of $B^0$ ($\bar{B}^0$) [(a)], and $B^{*0}$ ($\bar{B}^{*0}$) [(b)] are plotted as functions of magnetic field ($|eB|$) in units of $m_{\pi}^2$, at zero density $\rho_B = 0$, taking into account the effects of magnetized DS. The spin mixing between $B^0-B^{||*0}$ and $\bar{B}^0-\bar{B}^{||*0}$ are considered, as well as the effects of nucleons' AMMs. Comparison is made when these effects are not taken into account. } 
    \label{fig:1}
\end{figure}
\begin{figure}[h!]
    \centering
\includegraphics[width=1.0\textwidth]{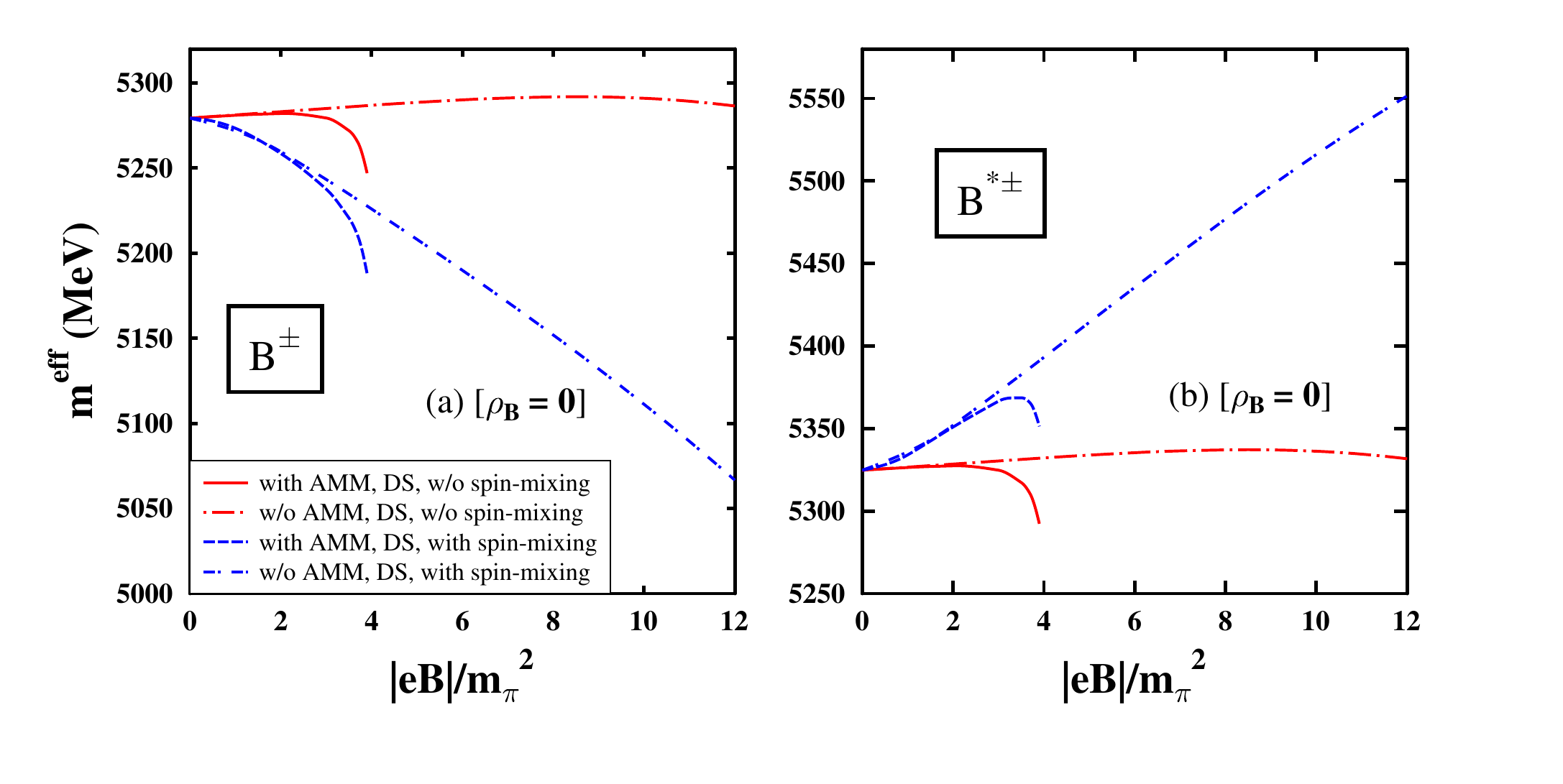}
    \vspace{-2.5em}
    \caption{In-medium masses of $B^{\pm}$ [(a)] and $B^{*\pm}$ [(b)] are plotted as functions of magnetic field in units of $m_{\pi}^2$, at zero density $\rho_B = 0$, accounting for the Dirac sea (DS) effects. The spin mixing effects between $B^{\pm}-B^{||*\pm}$ are considered, as well as the anomalous magnetic moments of the nucleons. Contributions of the lowest Landau level (LLL) are considered for the charged particles. Comparison is made with the case when the effects
    of spin-mixing and AMMs are not taken into account. } 
    \label{fig:2}
\end{figure}
In figure \ref{fig:1}, the effective masses of the neutral open bottom mesons, $B^0$, $\bar{B}^0$, $B^{*0}$ and $\bar{B}^{*0}$, are plotted as functions of $|eB|/m_{\pi}^2$, at $\rho_B=0$, by taking into account the effects of the Dirac sea and the AMMs of the nucleons in presence of an external magnetic field. As the solutions of fields are obtained till $|eB|=3.9 \ m_{\pi}^2$ for non-zero nucleons' AMMs, masses are shown around $|eB|\approx 4 m_{\pi}^2$ in this case. At $\rho_B=0$, there is striking difference between the two cases of with AMMs and without AMMs through the Dirac sea effect. The masses are plotted accounting for the spin magnetic field interaction between the pseudoscalar and the longitudinal component of the vector mesons. Thus, accounting for the spin- mixing between $(B^0-B^{||*0})$ and $\bar{B}^0-\bar{B}^{||*0}$, for the neutral open bottom mesons, with no lowest Landau level contribution, there is observed to be a level repulsion between their masses, leading to a rise (drop) in the masses of $S_z=0$ component of the vector (pseudoscalar) meson states with increasing magnetic field. It is compared to the case when the mixing effect is not considered. The masses for the particle and antiparticle of $B^0$ as well as $B^{*0}$ are same as shown in figure \ref{fig:1}, which can be argued from their self energy expressions in the dispersion relation (\ref{dis}) of $B$ mesons, due to the vanishing protons' and neutrons' number densities ($\rho_{p,n}$) in the vacuum \cite{div91,nikhil}. In figure \ref{fig:2}, in-medium masses of the charged open bottom mesons, $B^+$, $B^-$, $B^{*+}$ and $B^{*-}$, are plotted as functions of $|eB|/m_{\pi}^2$, at $\rho_B=0$, by taking into account the effects of the magnetized Dirac sea of nucleons, in an external magnetic field. The medium effects studied are the same as in figure \ref{fig:1}. However, for the charged  mesons ($B^{\pm}$), an additional contribution of the lowest Landau level (LLL) at finite magnetic field is taken into account, in the spin-magnetic field interactions between $B^+-B^{||*+}$ and $B^--B^{||*-}$ meson states. 
There is an overall sharp decrease in the masses of the neutral $B$ mesons (without the mixing effect) with increasing magnetic field in figure \ref{fig:1}, due to the magnetized Dirac sea contribution. Whereas this decrement is resisted in figure \ref{fig:2}, for the charged $B$ mesons mass, as there is a positive mass shift due to the LLL contribution (without spin-mixing) at finite magnetic field. The spin-mixing effect lead to explicitly visible rise (drop) in the mass of the $V^{||}$ and $P$ states with increasing magnetic field. 
\begin{figure}[h!]
    \centering
    \includegraphics[width=1.0\textwidth]{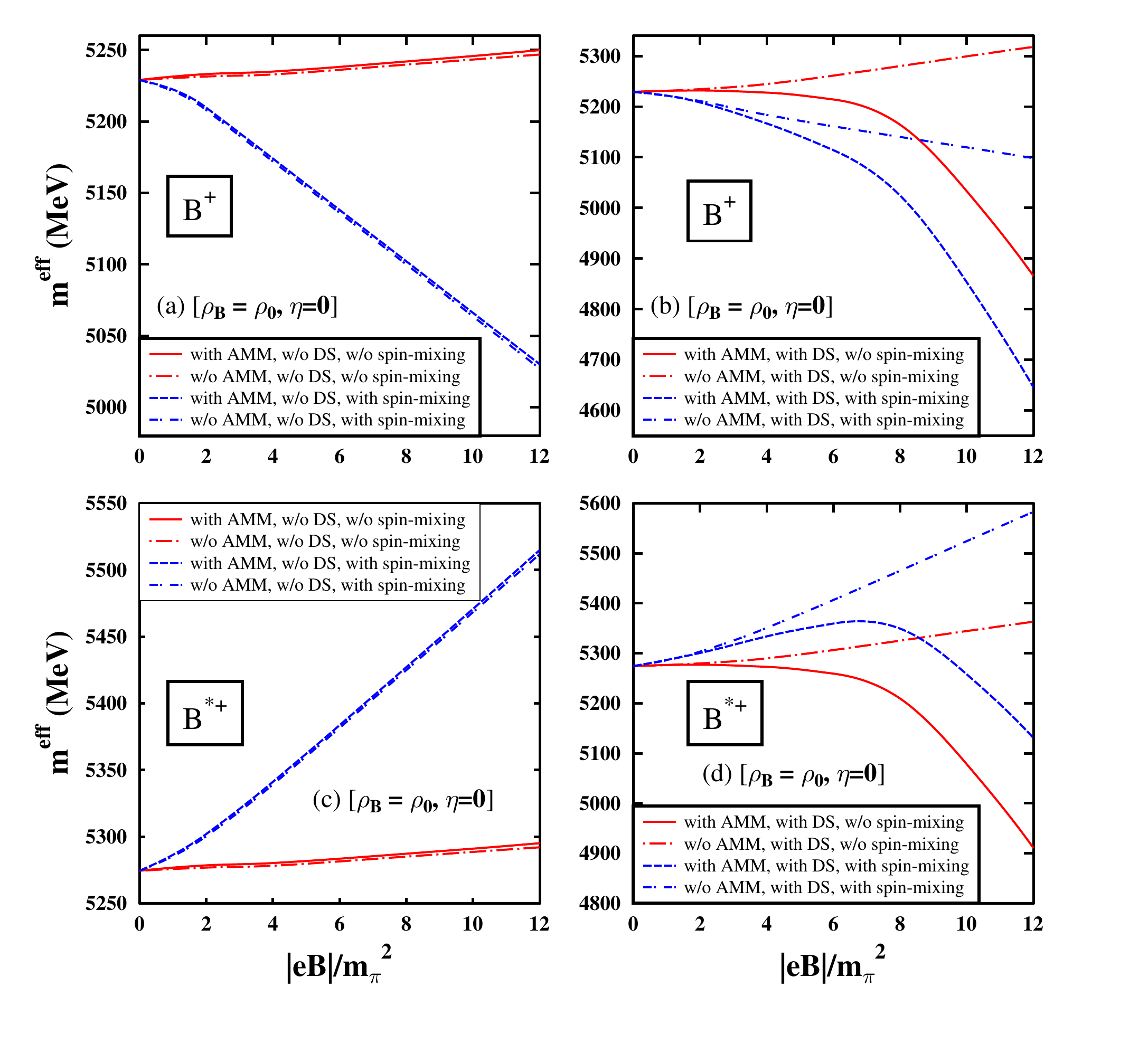}
    \vspace{-2.5em}
    \caption{In-medium masses of $B^+$ [(a)-(b)] and $B^{*+}$ [(c)-(d)] are plotted as functions of $|eB|/m_{\pi}^2$, at $\rho_B = \rho_0$, for symmetric nuclear matter, $\eta=0$, accounting for the Dirac sea (DS) effects. The spin mixing effects between $B^+-B^{||*+}$ are considered, along with the LLL contribution for the charged $B$ mesons. Effects of AMMs of the nucleons are compared with the no AMM case. Comparison is made with the case when the effects of spin-mixing and AMMs are not taken into
    account. } 
    \label{fig:3}
\end{figure}

\begin{figure}[h!]
    \centering
    \includegraphics[width=1.0\textwidth]{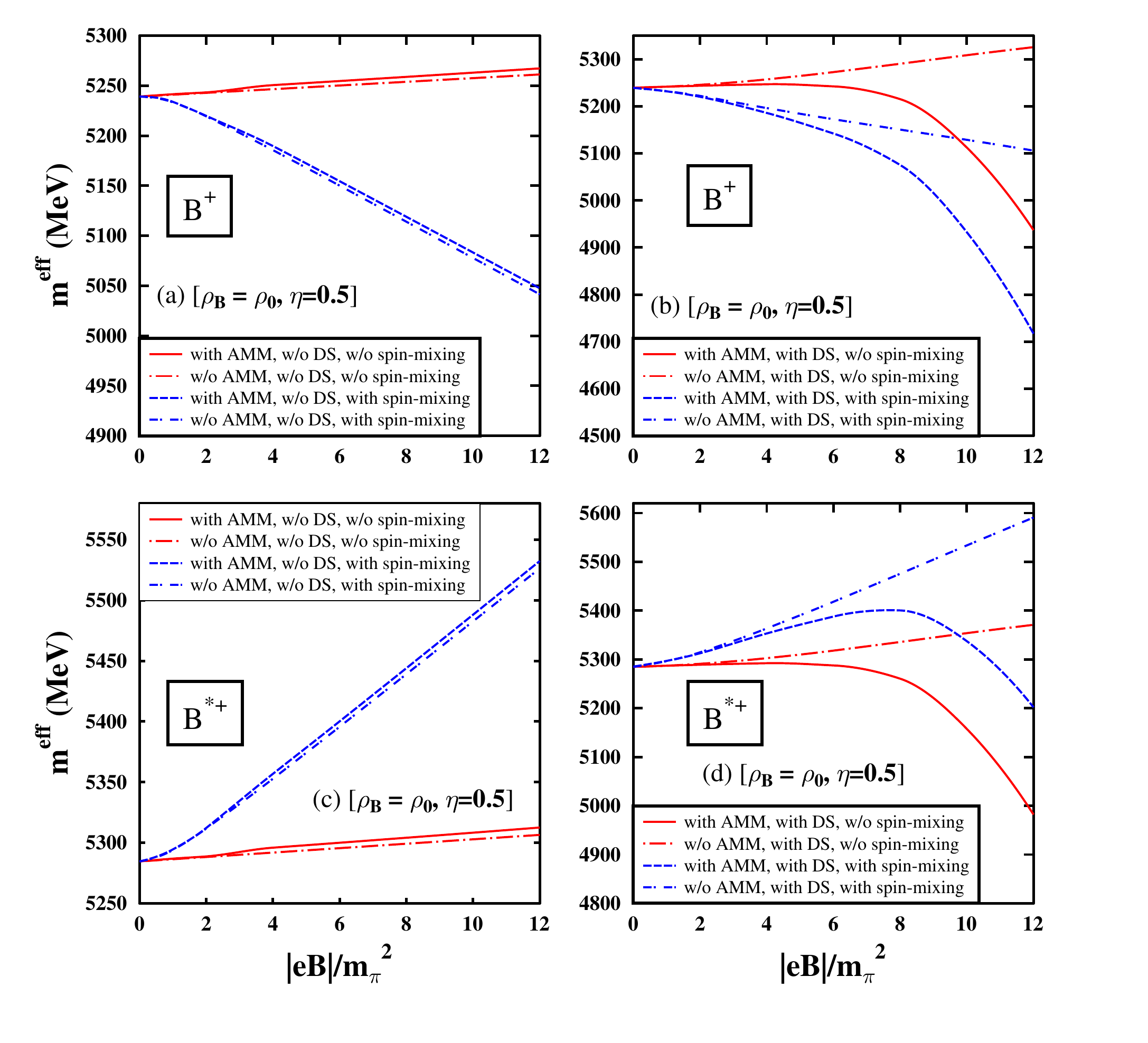}
    \vspace{-2.5em}
    \caption{In-medium masses of $B^+$ [(a)-(b)] and $B^{*+}$ [(c)-(d)] are plotted as functions of $|eB|/m_{\pi}^2$, at $\rho_B = \rho_0$, for asymmetric nuclear matter, $\eta=0.5$, accounting for the Dirac sea (DS) effects. Similar effects on the masses are shown as in figure \ref{fig:3}.} 
    \label{fig:4}
\end{figure}

\begin{figure}[h!]
    \centering
    \includegraphics[width=1.0\textwidth]{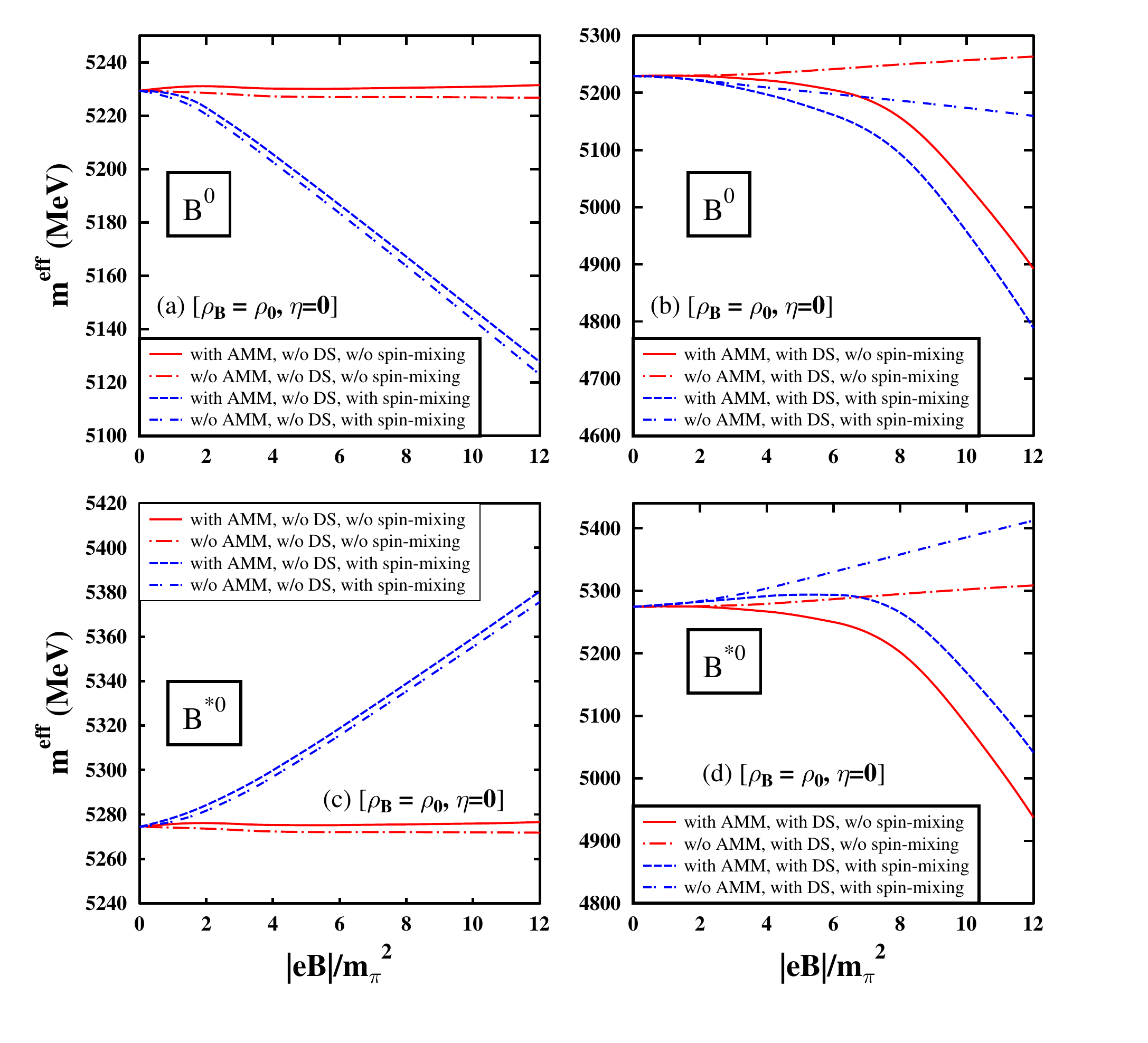}
    \vspace{-2.5em}
    \caption{In-medium masses of $B^0$ [in (a) and (b)] and $B^{*0}$ [in (c) and (d)] are plotted as functions of $|eB|/m_{\pi}^2$, at $\rho_B = \rho_0$, for symmetric nuclear matter, $\eta=0$, accounting for the Dirac sea (DS) effects. The spin mixing effects between $B^0-B^{||*0}$ are considered, with no LLL contribution for the charge neutral $B$ mesons. Effects of AMMs of the nucleons are compared with the no AMM case. Comparison is made when the effects of DS, spin-mixing, are not taken into account.} 
    \label{fig:5}
\end{figure}

\begin{figure}[h!]
    \centering
    \includegraphics[width=1.0\textwidth]{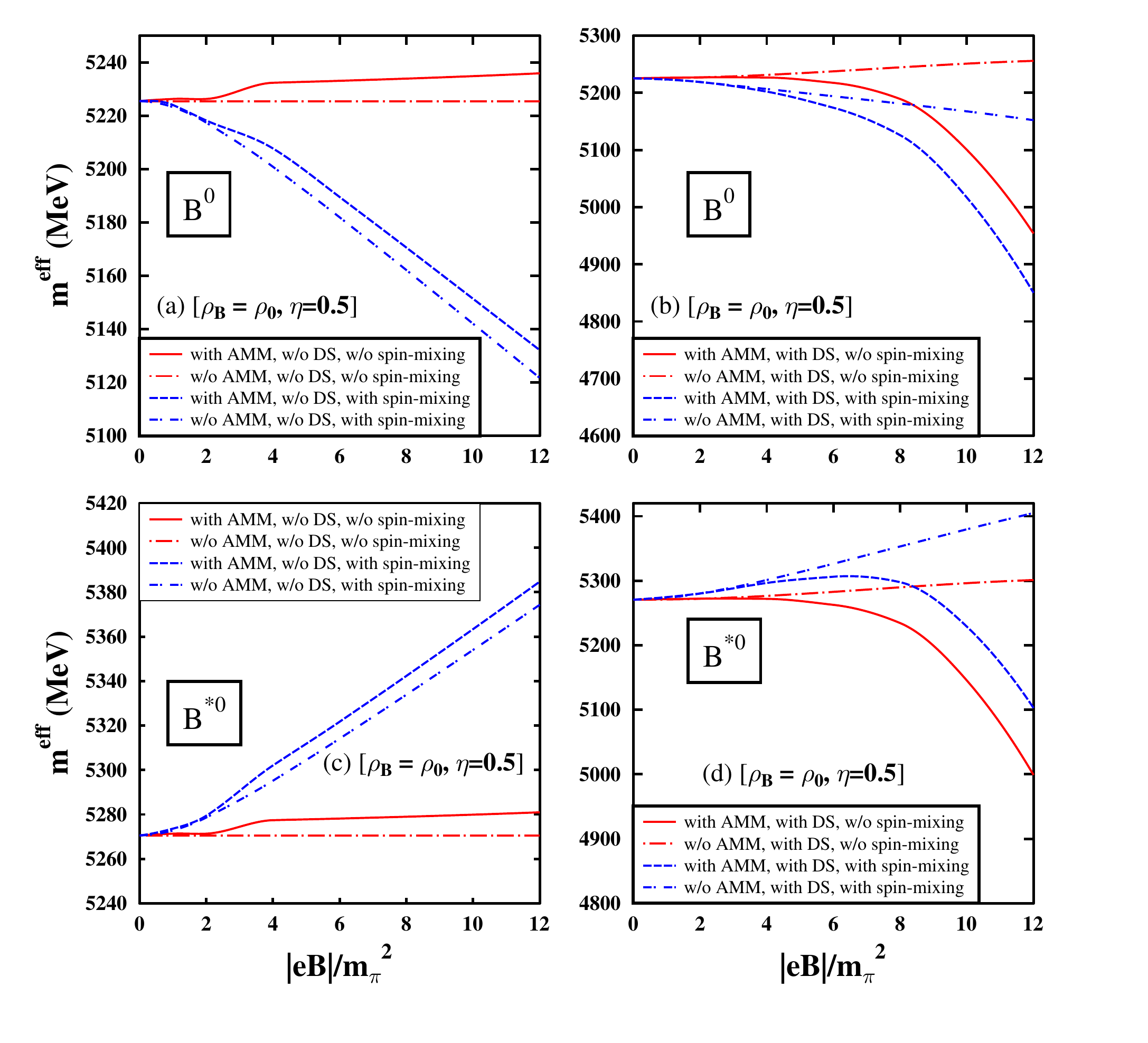}
    \vspace{-2.5em}
    \caption{In-medium masses of $B^0$ [in (a) and (b)] and $B^{*0}$ [in (c) and (d)] are plotted as functions of $|eB|/m_{\pi}^2$, at $\rho_B = \rho_0$, for asymmetric nuclear matter, $\eta=0.5$, accounting for the Dirac sea (DS) effects. Similar effects are studied as in fig.\ref{fig:5}.} 
    \label{fig:6}
\end{figure}
\begin{figure}[h!]
    \centering \includegraphics[width=1.0\textwidth]{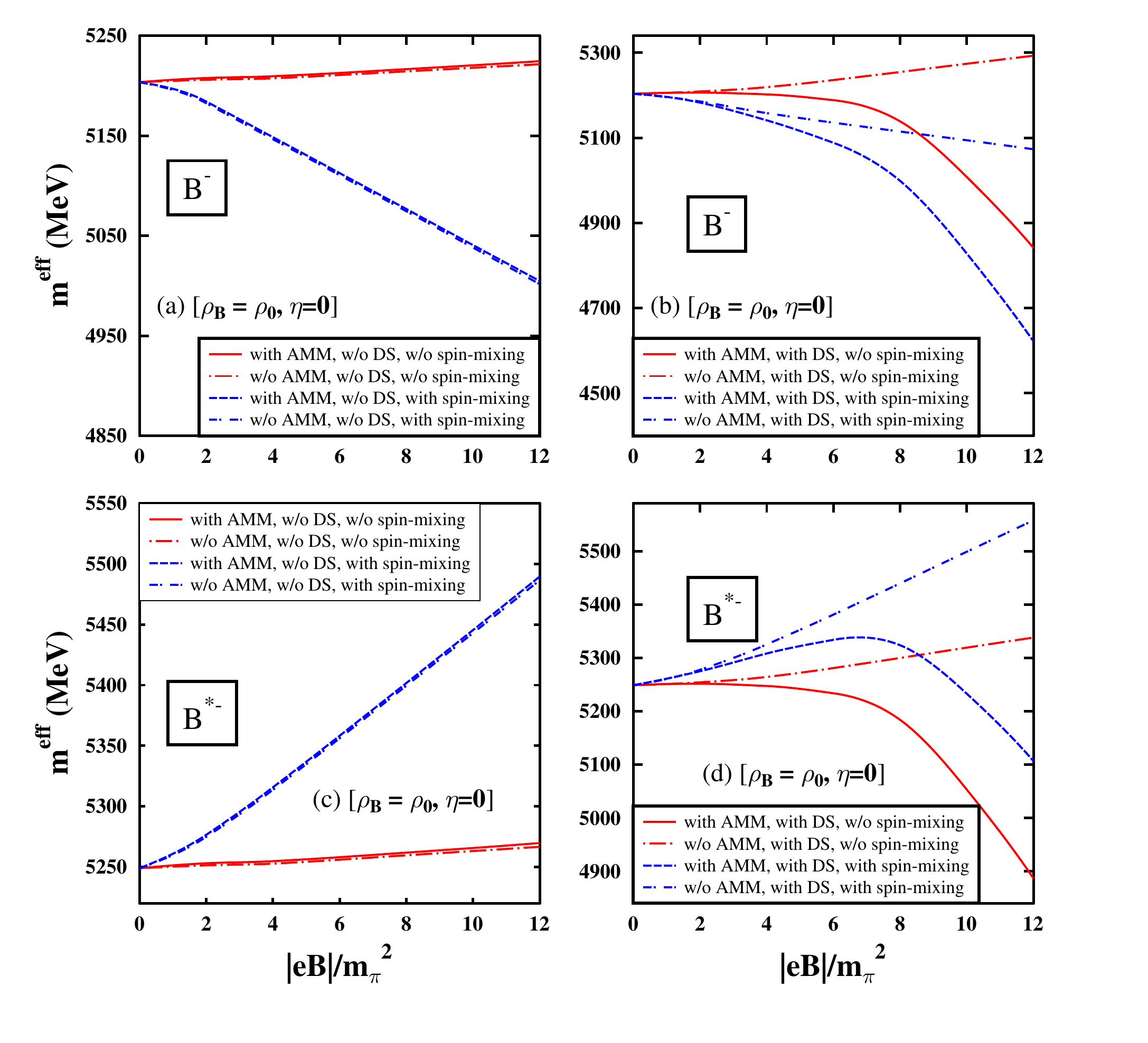}
    \vspace{-2.5em}
    \caption{In-medium masses of $B^-$ [in (a) and (b)] and $B^{*-}$ [in (c) and (d)] are plotted as functions of $|eB|/m_{\pi}^2$, at $\rho_B = \rho_0$, for symmetric nuclear matter, $\eta=0$, accounting for the Dirac sea (DS) effects. The spin mixing effects between $(B^--B^{||*-})$ are considered, along with the LLL contribution for the charged $B$ mesons. Effects of AMMs of the nucleons are compared with the no AMM case. Comparison is made when the effects of DS, spin-mixing, are not considered.} 
    \label{fig:7}
\end{figure}
\begin{figure}[h!]
    \centering   \includegraphics[width=1.0\textwidth]{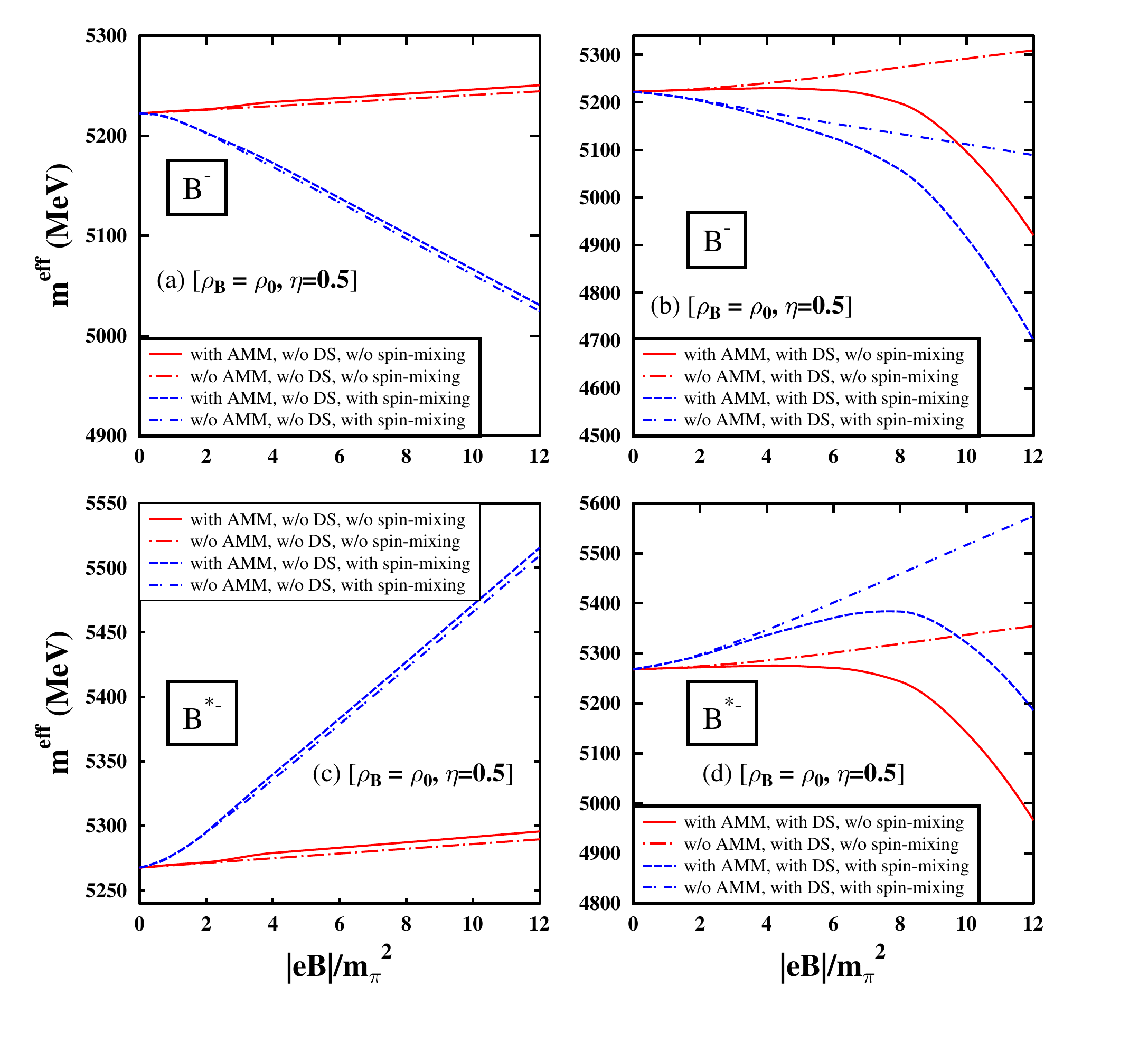}
    \vspace{-2.5em}
    \caption{In-medium masses of $B^-$ [in (a) and (b)] and $B^{*-}$ [in (c) and (d)] are plotted as functions of $|eB|/m_{\pi}^2$, at $\rho_B = \rho_0$, for the asymmetric nuclear matter, $\eta=0.5$, accounting for the Dirac sea (DS) effects. The in-medium effects studied on the masses are similar to fig.\ref{fig:7}} 
    \label{fig:8}
\end{figure}
\begin{figure}[h!]
    \centering \includegraphics[width=1.0\textwidth]{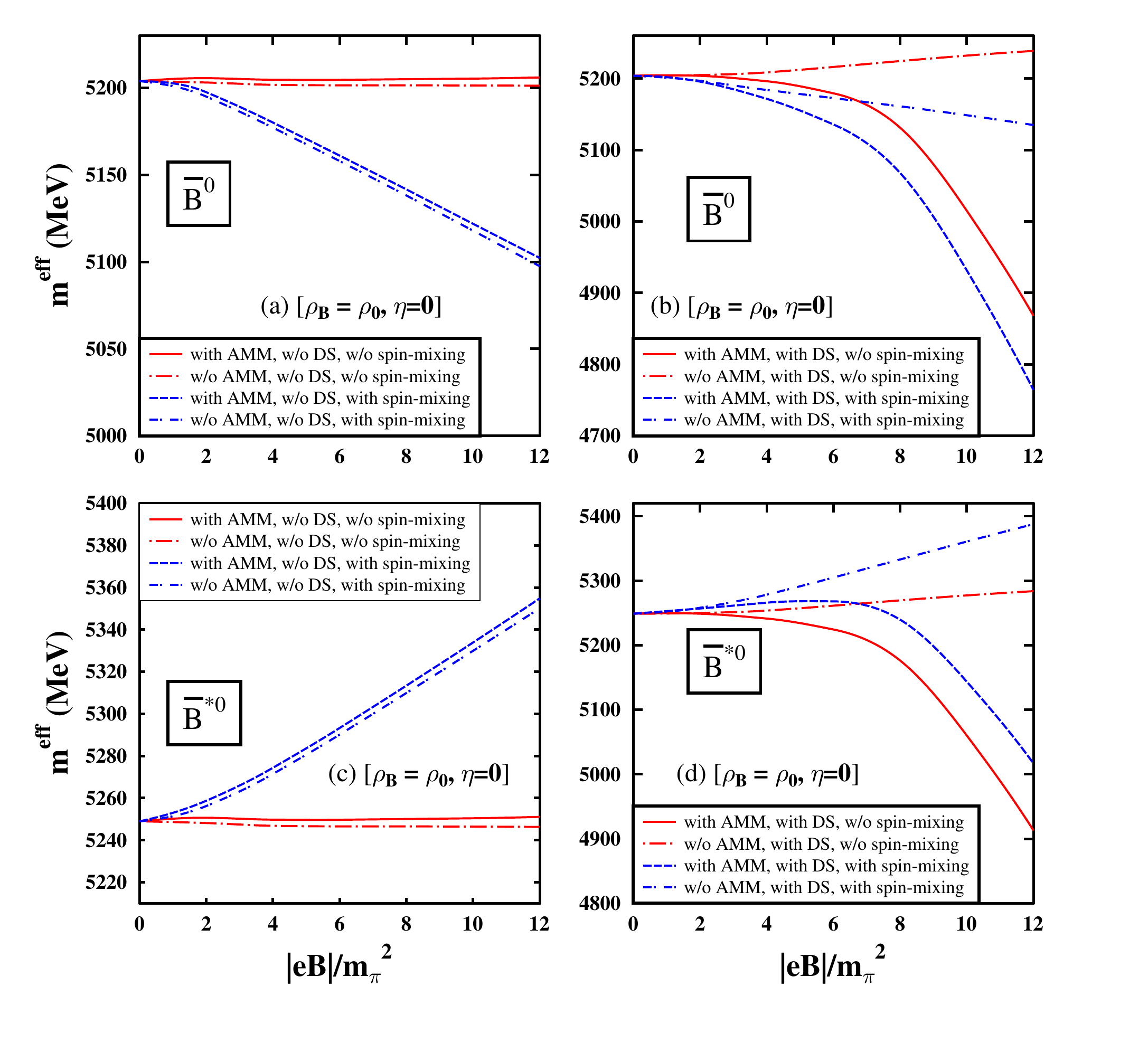}
    \vspace{-2.5em}
    \caption{In-medium masses of $\bar{B}^0$ [in (a) and (b)] and $\bar{B}^{*0}$ [in (c) and (d)] are plotted as functions of $|eB|/m_{\pi}^2$, at $\rho_B = \rho_0$, for symmetric nuclear matter, $\eta=0$, accounting for the Dirac sea (DS) effects. The spin mixing effects between $\bar{B}^0-\bar{B}^{||*0}$ are considered, but no LLL contribution for the charge neutral $B$ mesons. Effects of nucleonic AMMs are compared to the no AMM case. Comparison are made when the effects of DS, spin-mixing, are not taken into account.} 
    \label{fig:9}
\end{figure}
\begin{figure}[h!]
    \centering
\includegraphics[width=1.0\textwidth]{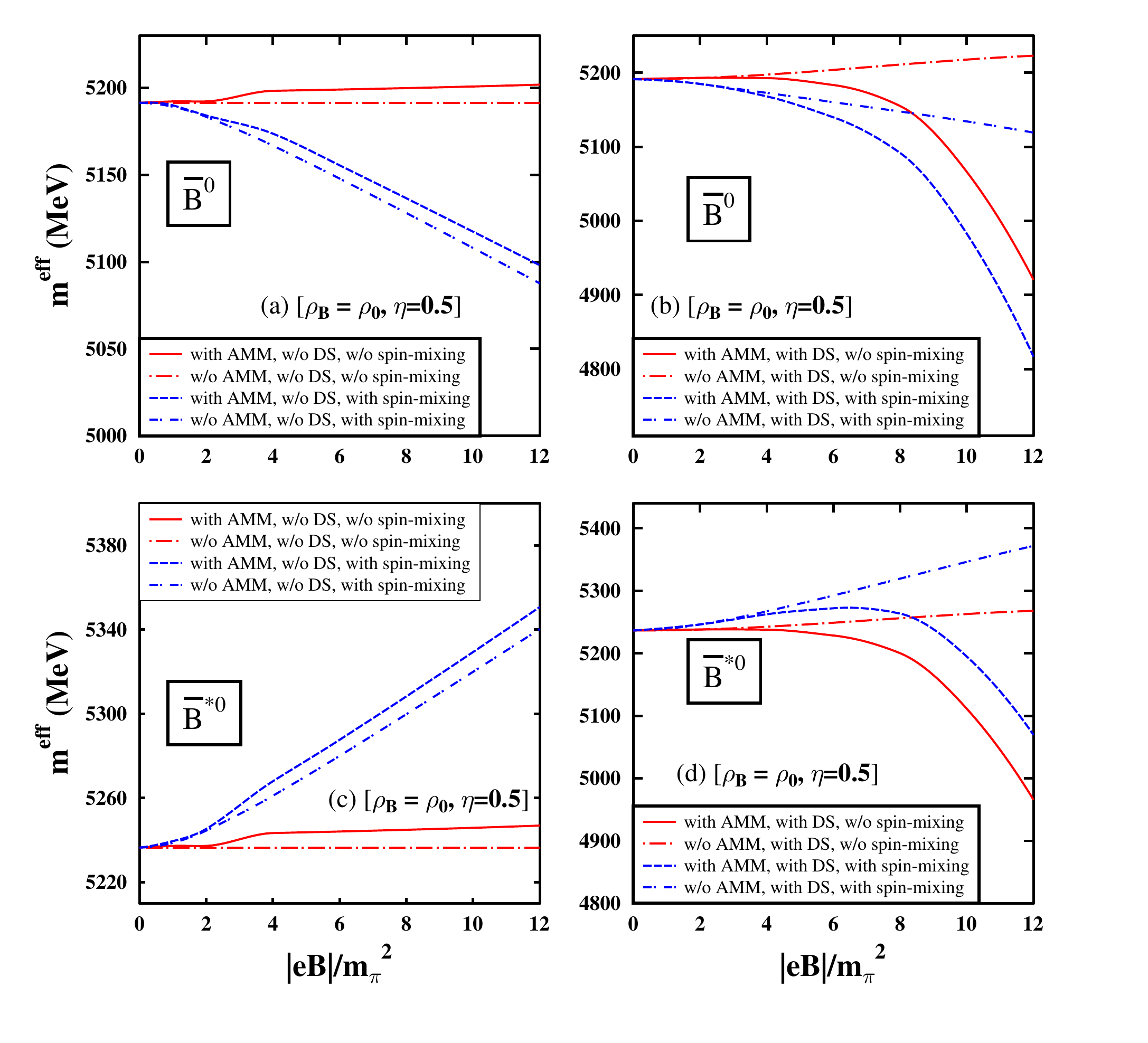}
    \vspace{-2.5em}
    \caption{In-medium masses of $\bar{B}^0$ [in (a) and (b)] and $\bar{B}^{*0}$ [in (c) and (d)] are plotted as functions of $|eB|/m_{\pi}^2$, at $\rho_B = \rho_0$, for asymmetric nuclear matter, $\eta=0.5$, accounting for the DS effects. Rest of the illustration are similar to fig. \ref{fig:9}} 
    \label{fig:10}
\end{figure}
\begin{figure}[h!]
    \centering
\includegraphics[width=1.0\textwidth]{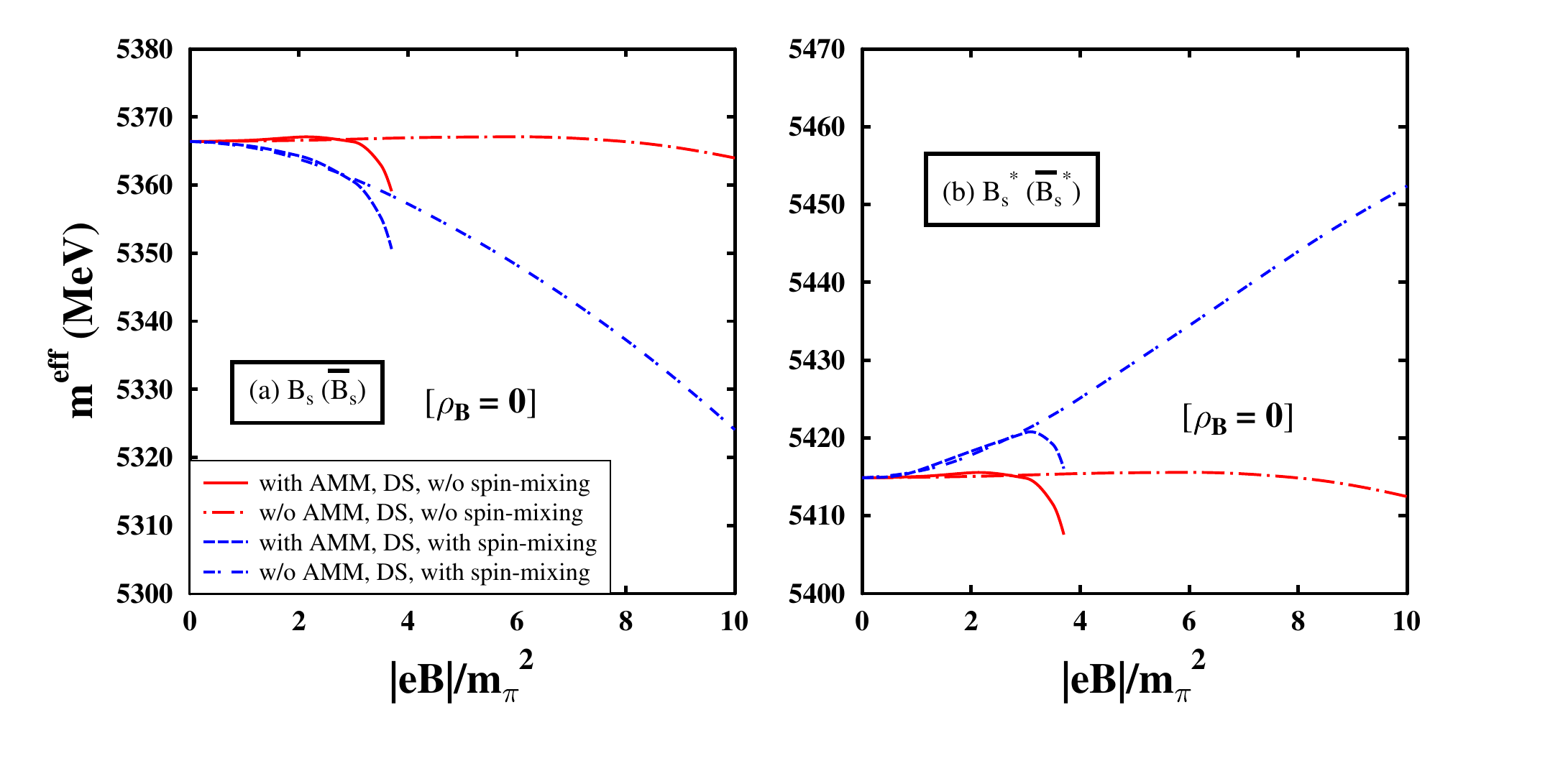}
    \vspace{-2.5em}
    \caption{In-medium masses of $B_s^0\ (\bar{B}_s^0)$ [in (a)] and $B_s^{*0}\ (\bar{B}_s^{*0})$  [in (b)] are plotted as functions of $|eB|/m_{\pi}^2$, at $\rho_B = 0$, accounting for the Dirac sea (DS) effects. The spin-magnetic field interaction between $B_s-B_s^{*||}$ and $\bar{B}_s-\bar{B}_s^{||*}$ lead to the mixing of the P and $V^{||}$ states, with no LLL contribution. The effects of nucleons' AMMs are compared to the no AMM case. Comparison is made when the effects of DS, spin-mixing, are not considered on the mass.} 
    \label{fig:11}
\end{figure}
\begin{figure}[h!]
    \centering
\includegraphics[width=1.0\textwidth]{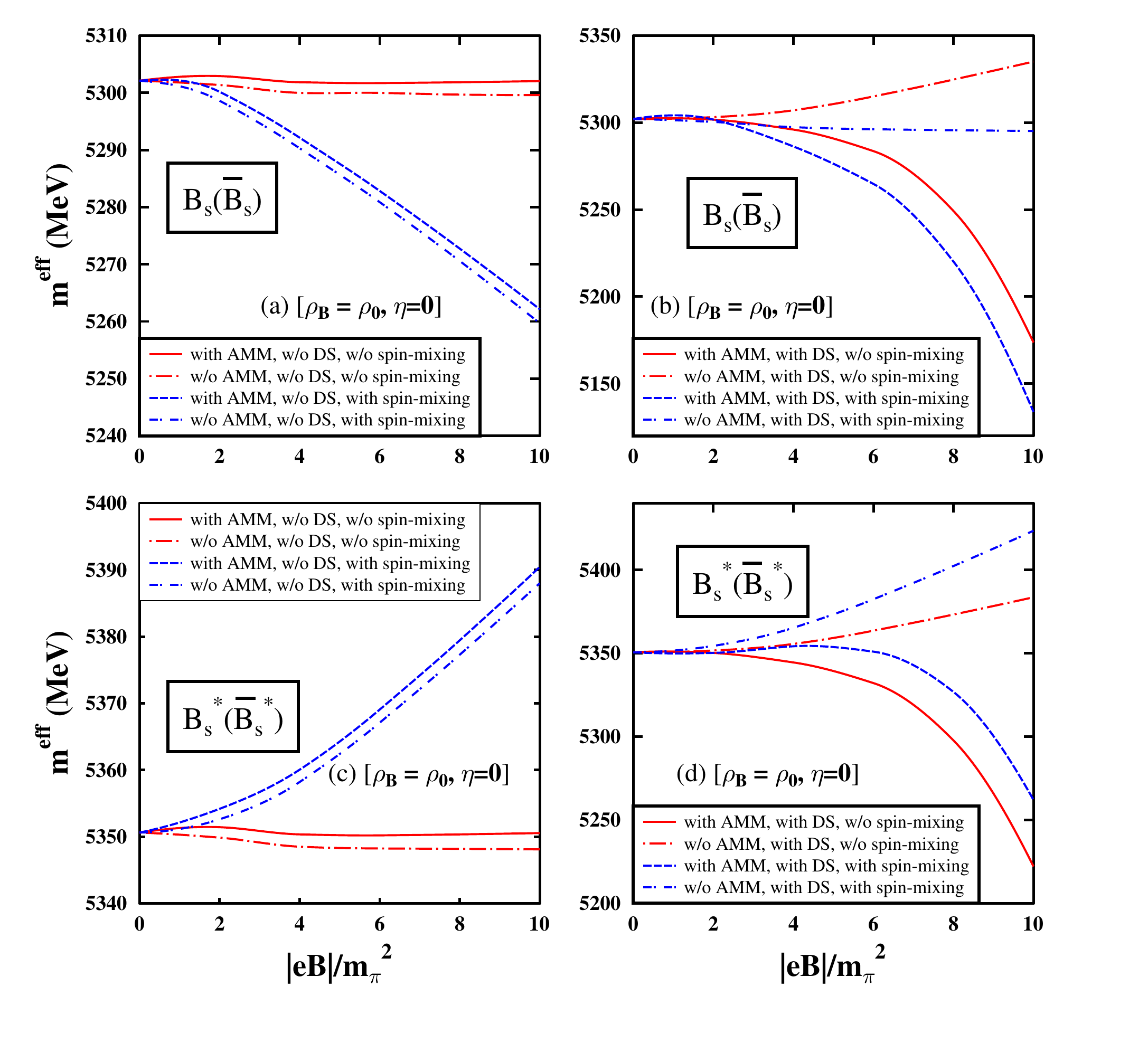}
    \vspace{-2.5em}
    \caption{In-medium masses of $B_s^0\ (\bar{B}_s^0)$ [in (a) and (b)] and $B_s^{*0}\ (\bar{B}_s^{*0})$ [in (c) and (d)] are plotted as functions of $|eB|/m_{\pi}^2$, at $\rho_B = \rho_0$, for asymmetric nuclear matter, $\eta=0$, accounting for the Dirac sea (DS) effects. The spin mixing effects between $B_s-B_s^{*||}$ and $\bar{B}_s-\bar{B}_s^{||*}$ are considered, without any LLL contribution. Rest of the effects studied are similar to fig. \ref{fig:11}.} 
    \label{fig:12}
\end{figure}

At the nuclear matter saturation density, $\rho_0$, for symmetric with $\eta=0$ (as well as the asymmetric with $\eta=0.5$) nuclear matter, the masses of the $B^+$-$B^{*+}$, $B^0$-$B^{*0}$, $B^{-}$-$B^{*-}$, and $\bar{B}$-$\bar{B}^{*0}$, are plotted as functions of $|eB|/m_{\pi}^2$, in figures \ref{fig:3}, \ref{fig:5}, \ref{fig:7}, \ref{fig:9} for the $\eta=0$ case and in figures \ref{fig:4}, \ref{fig:6}, \ref{fig:8}, \ref{fig:10} for the $\eta=0.5$ case, respectively, considering the Dirac sea effects. The contribution of the magnetic field is observed to be significant on the effective masses through the magnetized Dirac sea effects (in plots (b) and (d) of these figures) as compared to the no DS condition (in plots (a) and (c)), specifically on accounts of the anomalous magnetic moments of the nucleons. The spin-mixing effect lead to an increase (decrease) in the masses of $B^{||*+}(B^+)$, $B^{||*0}(B^0)$, $B^{||*-}(B^-)$ and $\bar{B}^{||*0}(\bar{B}^{0})$ meson states with magnetic field due to the corresponding mixing effects between $(B^+-B^{*+})$, $(B^0-B^{*0})$, $(B^--B^{*-})$ and $(\bar{B}^0-\bar{B}^{*0})$. The observed mass variation relative to the magnetic field is a combined effect of the magnetized Dirac sea, accounting for (and not) the anomalous magnetic moments of nucleons, the LLL contribution (for the charged mesons only) and the spin-magnetic field interaction effect. The Landau level contribution of protons in magnetized nuclear matter, as observed in the absence of Dirac sea effect (denoted as w/o DS) is very negligible on the open bottom meson masses, with almost no change as a function of the field. The cases of with and without nucleonic AMM through the Dirac and the Fermi sea of nucleons are compared in every plots, as well as the cases of with and without mixing. In the spin-mixing calculation for the charged open bottom mesons, the contribution of the lowest Landau energy level is considered in addition to the in-medium Dirac sea effects. In these figures, an important feature of magnetic field through the DS effect is observed that, for all the states overall mass decreases with increasing magnetic field at $\rho_B=\rho_0$ for ($\eta=0,\ 0.5$), and non zero AMMs of the nucleons, an effect of inverse magnetic catalysis as it is described in the beginning of the present section, in the absence of mixing effects. In the presence of mixing effect, due a positive ($V^{||}$) and negative ($P$) mass shifts there is seen to be a slightly lower rate in the mass drops relative to $eB$ for the vector mesons but for the pseudoscalar states, mass drops at a relatively higher rate than w/o spin-mixing case. On the other hand, for zero nucleonic AMM, the opposite behavior is obtained due to the effect of magnetic catalysis in this case. In-medium mass of the pseudoscalar and vector open bottom mesons increases with increasing magnetic field at the nuclear matter saturation density both in symmetric as well as asymmetric nuclear matter. This increment is affected in opposite way by the spin-mixing effects for the pseudoscalar and vector meson states, as can be seen from the figures. Thus, plots (a) and (c) in these figures indicate that the dominant contribution is coming due to the magnetic field induced mixing between the pseudoscalar and longitudinal component of vector meson states with negligible effects from the Landau energy levels of protons and anomalous magnetic moments of the nucleons in the magnetized Fermi sea of nucleons. This is not the situation when Dirac sea effect is accounted for. In plots (b) and (d) the dominant magnetic field effect is obtained through the magnetized DS specially for finite nucleonic AMMs. \\
As it is argued in section \ref{sec2}, the mass shifts of the vector open bottom mesons are the same with that of the pseudoscalar partners within the chiral effective model, given by equation (\ref{diff}). The vacuum mass of all four vector mesons $B^{*0}, \bar{B}^{*0}$ , $B^{*\pm}$ are taken to be $5324.71$ MeV, and that of the pseudoscalar mesons are taken to be $5279.34$ MeV for $B^{\pm}$ and $5279.66$ MeV for $B^0(\bar{B}^0)$ \cite{pdg}. From the mass shifts of the pseudoscalar mesons, using equation (\ref{diff}), the in-medium effective masses of the vector open bottom mesons are obtained at different values of the medium parameters from their pseudoscalar partners. In the plots for the charged vector particles, wherever the LLL contribution is considered it is for the longitudinal part only with $S_z=0$ in equation (\ref{lllvec}), as it is used in the spin-mixing calculations.\\ 
In the presence of an external magnetic field, the mixing phenomenon between the pseudoscalar and the
longitudinal component of the vector mesons have been studied using a phenomenological effective Lagrangian approach with the coupling parameter fitted from the empirical value of the radiative decay width of a vector meson going to a pseudoscalar and a photon \cite{102, amsm30}. This effect strongly modify the in-medium partial decay widths of $D^*\rightarrow D\pi$ as well as of the $\psi(3770)\rightarrow D\bar{D}$ due to the contributions of the $\psi(3770)-\eta'_c$ mixing to the mass of the charmonium state $\psi(3770)$, as well as due to the $(D-{D}^{*})$ and $(\bar{D}-\bar{D}^{*}$) mixing contributions to the masses of the open charm
mesons ($D\bar{D}$) \cite{102, amsm30}. Due to the lack of experimental data on the radiative decays ($V\rightarrow P\gamma$) in the bottom sector, an interaction Hamiltonian approach is adopted to account for the spin-magnetic field interaction on the open bottom mesons, as given by equation (\ref{hamil}). In ref.\cite{amsm31}, an interaction Hamiltonian approach was used to study the spin-mixing effects on the $\Upsilon(4S)-\eta_b(4S)$, $B-{B}^{*}$ and $\bar{B}-\bar{B}^{*}$ mesons, with the masses calculated within the chiral effective model, accounting for the magnetized Fermi sea effects and lowest Landau level contribution (for $B^{\pm}$) \cite{amsm31}. The magnetic field effects through this mixing have been observed to modify considerably the in-medium partial decay widths of $\Upsilon(4S)\rightarrow B\bar{B}$, calculated using a field theoretical model of composite hadrons with quark (and antiquark) constituents \cite{amsm31}.\\
The in-medium masses of the pseudoscalar bottom strange mesons $B_s^{0}$, $\Bar{B}_s^{0}$ and corresponding vector meson states $B_s^{*0}$, $\Bar{B}_s^{*0}$ are shown relative to the variation in magnetic field $|eB|/m_{\pi}^2$, in figure \ref{fig:11} at vacuum and in figure \ref{fig:12} at $\rho_B=\rho_0$, symmetric nuclear matter, accounting for the effects of magnetized Dirac sea. At non zero magnetic field, there is no LLL contribution to the masses due to the charge neutrality of such mesons. Effects of spin-magnetic field interaction between $B_s-B_s^{*}$ (their antiparticle states) are considered on the mass modifications. The mass of the particle and antiparticle states of $B_s$ mesons are same in nuclear matter which can be inferred from the self energy expression of the $B_s$ mesons given by equation (\ref{selfBs}). This fact is also evident from the plots of mass variation of $B_s$ mesons with respect to magnetic field in figures \ref{fig:11}-\ref{fig:12}. \\   
Thus, the study of in-medium masses of the open bottom mesons accounting for the effects of Dirac sea in magnetized nuclear matter, show important mass modifications of the open bottom mesons due to the (inverse) magnetic catalysis. This can significantly contribute to the in-medium partial decay widths of $\Upsilon(4S)\rightarrow B\Bar{B}$ due to the magnetized Dirac sea contribution, which may further lead to the modifications in the yields of the open bottom and bottomonia in the peripheral ultra relativistic heavy ion collision experiments where produced magnetic field is huge.

\section{Summary}
\label{sec4}
The in-medium masses of the pseudoscalar ($B^0,\ \bar{B}^0,\ B^{+}, B^{-}$, $B_s^0$, $\Bar{B}_s^{0}$) and vector ($B^{*0},\ \bar{B}^{*0},\ B^{*+}, B^{*-}$, $B_s^{*0}$, $\Bar{B}_s^{*0}$) open bottom mesons are studied in the magnetized (asymmetric) nuclear matter, incorporating the effects of magnetized Dirac sea, within the chiral effective model framework. The contribution of the lowest Landau level (LLL) are considered by neglecting the internal structure of the charged meson particles. The (reduction) enhancement in the light quark condensates with increasing magnetic field lead to the phenomena of (inverse) magnetic catalysis due to the contribution of magnetic fields on the Dirac sea. There is seen to be significant modifications on the masses accounting for the effects of (inverse) magnetic catalysis as compared to the no sea effects in the magnetized nuclear matter. In the vacuum, the magnetic fields are seen to have an appreciable contribution on the masses (apart from the LLL for $B^{\pm}$), due to the Dirac sea. The anomalous magnetic moments of the nucleons are seen to have important effects at zero as well as the nuclear matter saturation density via Dirac sea contribution. The pseudoscalar and (longitudinal component of) vector mesons mixing including the Dirac sea effects are studied, which lead to a level repulsion between the masses of the longitudinal component of the vector and the pseudoscalar mesons, which increases with rising magnetic field. The visibly large effects of the (inverse) magnetic catalysis on the masses of the open bottom mesons at finite magnetic field should have important observable consequences in the non central, high energy heavy-ion collision experiments where produced magnetic fields are estimated to be large.

\end{document}